\documentclass[12pt,onecolumn]{IEEEtran}
\usepackage{amsmath,amssymb,euscript ,yfonts,psfrag,latexsym,dsfont,graphicx,bbm,color,amstext,wasysym,subfig,flushend,parskip,textcomp,cite,caption}
\graphicspath{{./},{./figures/}}

\begin{document}
\newtheorem{thm}{Theorem}
\newtheorem{cor}[thm]{Corollary}
\newtheorem{conj}[thm]{Conjecture}
\newtheorem{lemma}[thm]{Lemma}
\newtheorem{prop}{Proposition}
\newtheorem{problem}[thm]{Problem}
\newtheorem{remark}[thm]{Remark}
\newtheorem{defn}[thm]{Definition}
\newtheorem{ex}[thm]{Example}

\newcommand{\mR}{{\mathbb R}}
\newcommand{\cR}{{\mathcal R}}
\newcommand{\f}{{\mathfrak f}}

\newcommand{\diag}{\operatorname{diag}}
\newcommand{\trace}{\operatorname{trace}}
\newcommand{\argmin}[1]{\underset{#1}{\operatorname{arg}\operatorname{min}}\;}
\newcommand{\degree}{{}^\circ}
\newcommand{\ignore}[1]{}

\def\spacingset#1{\def\baselinestretch{#1}\small\normalsize}
\setlength{\parskip}{2pt}
\setlength{\parindent}{10pt}
\setlength{\textfloatsep}{8pt plus 1.0pt minus 2.0pt}
\spacingset{1}

\newcommand{\mike}{\color{magenta}}
\definecolor{grey}{rgb}{0.6,0.6,0.6}
\definecolor{lightgray}{rgb}{0.97,.99,0.99}
\captionsetup[figure]{font=small,labelfont=small}

\title{Rotated spectral principal component analysis (rsPCA) for identifying dynamical modes of variability in climate systems
}

\author{Cl\'ement~Guilloteau, Antonios~Mamalakis, Lawrence~Vulis, Tryphon~T.~Georgiou, and 
Efi~Foufoula-Georgiou
\thanks{C. Guilloteau (cguillot@uci.edu), A. Mamalakis, L. Vulis, are with the Department of Civil and Environmental Engineering, University of California, Irvine}
\thanks{T. T. Georgiou is with the department of Mechanical and Aerospace Engineering, University of California, Irvine}
\thanks{E. Foufoula-Georgiou is with the Department of Civil and Environmental Engineering and the Department of Earth System Science, University of California Irvine}}

\markboth{\today}{}

\maketitle
\begin{abstract}
Spectral PCA (sPCA), in contrast to classical PCA, offers the advantage of identifying organized spatio-temporal patterns within specific frequency bands and extracting dynamical modes. However, the unavoidable tradeoff between frequency resolution and robustness of the PCs leads to high sensitivity to noise and overfitting, which limits the interpretation of the sPCA results. We propose herein a simple non-parametric implementation of the sPCA using the continuous analytic Morlet wavelet as a robust estimator of the cross-spectral matrices with good frequency resolution. To improve the interpretability of the results when several modes of similar amplitude exist within the same frequency band, we propose a rotation of eigenvectors that optimizes the spatial smoothness in the phase domain. The developed method, called rotated spectral PCA (rsPCA), is tested on synthetic data simulating propagating waves and shows impressive performance even with high levels of noise in the data. Applied to historical sea surface temperature (SST) time series over the Pacific Ocean, the method accurately captures the El Ni\~{n}o-Southern Oscillation (ENSO) at low frequency (2 to 7 years periodicity). At high frequencies (sub-annual periodicity), at which several extratropical patterns of similar amplitude are identified, the rsPCA successfully unmixes the underlying modes, revealing spatially coherent patterns with robust propagation dynamics. Identification of higher frequency space-time climate modes holds promise for seasonal to subseasonal prediction and for diagnostic analysis of climate models.
\end{abstract}


%
%
%
%
%
%
\section{Introduction}

Identifying spatio-temporal relations and defining coherent modes of variability across different regions and variables is the essence of characterizing complex systems. The Earth's climate system is obviously a complex system of particular interest, exhibiting a plethora of dynamic modes of variability originating from different physical processes (e.g. solar forcing, oceanic/atmospheric circulations, land/atmosphere interactions, etc.), and imprinting themselves at various spatial and temporal scales. The accurate identification and modelling of the modes of the climate system is necessary for many key problems in geosciences, such as weather/climate prediction, attribution of extreme events and hazards, and assessment of climate change impacts. 

The Earth's climate system includes the oceans, with sea surface temperature (SST) in particular being a variable whose space-time variability over a range of scales encapsulates the complex oceanic processes and their coupling with the atmosphere \cite{deser2010sea}. The SST is commonly used as indicator of multi-decadal trends and global warming \cite{trenberth2013apparent,douville2015recent,ferster2018recent} and also as a medium-range and seasonal predictor of multiple regional climatic variables such as temperature and precipitation\cite{ting1997summertime,uvo1998relationships,ali2013relationship,mckinnon2016long,mamalakis2018new} Because of the large extent of the ocean and the substantial variability of the SST, a large number of variables is necessary to comprehensively describe its spatio-temporal dynamics. To make such a large dataset interpretable and usable for diagnostic and prediction purposes, it is often necessary to find a proper dimensionality reduction scheme to reduce the number of variables of this complex system while minimizing loss of information; in other words, reduce the system to a manageable number of dynamical modes. 

A wide range of dimensionality reduction methods exists. Principal Component Analysis (PCA), also commonly referred to as Empirical Orthogonal Function (EOF) analysis, is a non-parametric method widely used in climate science \cite{Jolliffe1986,keiner1997empirical,hannachi2007empirical,navarra2010guide}. It consists in performing a Singular Value Decomposition (SVD) of a multivariate dataset through the computation of the covariance matrix between all the variables and extraction of its eigenvectors. Its simplicity of implementation as well as the fact that this empirical method does not require any a-priori assumption or model/parameter selection have contributed to its popularity and success. Nevertheless, the method has several limitations which are further discussed below. Many variations or extensions of PCA have been proposed to overcome some of these limitations. For example, if the variables are given under the form of regularly sampled time series, one can compute the lagged covariances or the Fourier cross-spectra between the variables rather than computing only the (zero-lag) covariances, and then construct the empirical lagged-covariance matrices or the cross-spectral matrices of the system. This gives rise to the lagged PCA and the spectral PCA as "natural" extensions of the classical PCA.

The lagged PCA allows to better extract dynamical modes when different variables or areas of the studied domain have delayed linear responses to the same signal with different delays. The spectral PCA (sPCA), through the phase information in the complex cross-spectral coefficients, also allows to handle lagged correlations. Additionally, it offers the possibility to look for modes in specific frequency bands and is particularly potent at extracting wave-type modes and to handle propagation effects (non-stationary waves). While sPCA (also known as frequency domain EOF) was introduced and theorized in the early $\mathrm{70's}$ \cite{wallace1972aempirical,wallace1972bempirical} and later re-defined and implemented under various forms \cite{horel1984complex,hasselmann1988pips,johnson1993structure,mann1994global,mann1999oscillatory,ghil2002advanced,thornhill2002spectral,mann2020absence}, it has not become a standard method in atmospheric and climate science. This may be due to the fact that, in spite of being relatively straightforward in theory, its implementation requires choosing an appropriate method for computing robust cross-spectral coefficients from finite-length time series and avoid overfitting. Additionally, as already pointed out by \cite{wallace1971spectral} and \cite{wallace1972aempirical}, the interpretability of the extracted Principal Components (PCs) may be difficult when several PCs of similar amplitude co-exist within the same spectral band.

To overcome the above difficulties, we herein propose a wavelet-based implementation of the sPCA which relies on the complex Morlet wavelet for the estimation of the cross-spectral matrices. The continuous wavelet transform is nowadays a standard and popular tool for spectral and cross-spectral analysis \cite{hudgins1993wavelet,perrier1995wavelet,kumar1997wavelet,jiang2011wavelet,banskota2017continuous}. The Morlet analytic wavelet in particular allows robust estimation of power spectra and cross-spectra in the frequency domain, even from relatively short time series, while allowing for reasonably good frequency localization \cite{kirby2005wavelet,cottis2016relationship}. In addition, we propose a rotation of the eigenvectors resulting from the wavelet-based sPCA, using the spatial regularity (smoothness) of the phase of the rotated vectors as an optimality criterion to select the "best" rotation. The combination of these two innovations (robust spectral estimation through the Morlet wavelet and phase regularization of the rotated PCs) improves the interpretability and reduces the sensitivity of the extracted modes to noise and small variations in the data.

The article is organized as follows. Section 2 starts with the classical PCA, introduces the implementation of sPCA through the complex Morlet wavelet transform, and describes the methodology for the rotation of the eigenvectors (rsPCA). In section 3, the proposed methodology is tested on synthetic data, namely numerically generated non-stationary waves propagating in a $\mathrm{2D}$ plane, plus random noise. Section 4 presents the results of applying the rsPCA method to historical $\mathrm{(1940-2018)}$ fields of SST over the Pacific Ocean, with a particular focus on identifying and interpreting the sub-annual modes of variability which have received relatively little attention in the climate literature. Discussion and conclusions are presented in section 5.
%
%
%
%
%
%
\section{Methodology}
Let us consider a dataset made of $\mathit{L}$ observations of $\mathit{N}$ variables (e.g., time series of length $\mathit{L}$ associated with $\mathit{N}$ spatial locations). This dataset corresponds to the $\mathit{N}$ by $\mathit{L}$ data matrix $\mathbf{X}$:
\begin{equation}
 \mathbf{X}=\left[\begin{matrix}
              x_{1,1}&\cdots&
              x_{1,L}\\\vdots&\ddots&\vdots\\
              x_{N,1}&\cdots&
              x_{N,L}
 \\\end{matrix}\right]
\end{equation}
In the sequel, the notation $\mathbf{x}_\mathit{i}=(x_{\mathit{i},1}, \ldots, x_{\mathit{i,L}})$, designates the $\mathit{i}^{th}$ row of the matrix, i.e. the vector of observations of the $\mathit{i}^{th}$ variable.
%
%
%
%
%
%
\subsection{Classical PCA}
The empirical sample covariance matrix $\mathbf{C}$ of the $\mathbf{X}$ dataset is a $\mathit{N}$ by $\mathit{N}$ matrix defined as:

\begin{equation}
        \mathbf{C}=\frac{1}{N}\mathbf{XX}^\prime
\end{equation}

with $\mathbf{X}^\prime$ denoting the transpose conjugate matrix of $\mathbf{X}$.
The principal component analysis is performed by extracting the eigenvectors $\mathbf{u}_\mathbf{i}$ and associated eigenvalues ${\lambda_i}^2$ of the covariance matrix, which are the solutions of the system:
\begin{equation}
\mathbf{Cu_i}={\lambda_\mathit{i}}^2\mathbf{u_i}
\end{equation}
The principal component time series associated with the eigenvector $\mathbf{u_i}$ is obtained as:
\begin{equation}
\mathrm{\boldsymbol{\kappa}_\mathbf{i}} = \mathbf{Xu_i}
\end{equation}
The rank of a system is the number of linearly independent variables composing it. The empirical rank of a system is taken as the rank of its covariance matrix, i.e. the number of linearly independent columns of the $\mathbf{C}$ matrix, which is equal to $\mathit{N}$ minus the number of linearly independent eigenvectors associated with the zero eigenvalue. A system of rank $\mathit{R}<\mathit{N}$, can be reduced to $\mathit{R}$ variables without loss of information. In practice, the covariance matrix derived from the data is generally found to be full rank (of rank $\mathit{N}$), simply because the relations between the variables of the system are not perfectly linear or because of measurement noise. However, PCs associated with small eigenvalues may be neglected with minimum loss of information. Therefore, the decay rate of the eigenvalues can be used as a measure of the "reducibility" of a system (i.e. how easily a high dimensional system can be compressed into a low number of modes without important loss of information). 

PCA is a powerful tool for extracting linear modes of variability of complex systems and has seen many applications in atmospheric sciences and beyond. However, it has one main limitation in that it can only identify linear "synchronous" relationships between the variables of a system. In particular, when several variables of the system have a delayed linear response to a given signal and if the time delay is not identical for all the variables, the PCA will generally fail to identify this dynamic relationship or will require several PCs to compressively capture it. 
%
%
%
%
%
%
\subsection{Spectral PCA (sPCA) via the Morlet wavelet}
Spectral PCA (sPCA) relies on the computation of the cross-spectral matrix $\mathbf{S_k}$ and extraction of its eigenvectors in various frequency bands $\mathit{b_k(f)}$. In the frequency band $\mathit{b_k(f)}$, the complex $\mathit{N}$ by $\mathit{N}$ matrix $\mathbf{S_k	}$ is defined as:
\begin{equation}
\mathbf{S_k}=\frac{1}{\int_{-\infty}^{+\infty}{b_k\left(f\right)df}}\\ 
\left[\begin{matrix}
     \int_{-\infty}^{+\infty}{s_{\mathit{x}_1,\mathit{x}_1}\left(f\right)b_k\left(f\right)df}&\cdots&
      \int_{-\infty}^{+\infty}{s_{\mathit{x}_1,\mathit{x}_N}\left(f\right)b_k\left(f\right)df}\\\vdots&\ddots&\vdots\\
      \int_{-\infty}^{+\infty}{s_{\mathit{x}_N,\mathit{x}_1}\left(f\right)b_k\left(f\right)df}&\cdots&
      \int_{-\infty}^{+\infty}{s_{\mathit{x}_N,\mathit{x}_N}\left(f\right)b_k\left(f\right)df}\\
      \end{matrix}\right]
\end{equation}
with $\mathit{s_{\mathbf{x}_i,\mathbf{x}_j}\left(f\right)}$ the Fourier cross-spectrum between the two time series $\mathbf{x}_\mathit{i}$ and $\mathbf{x}_\mathit{j}$ and $\mathit{b_k\left(f\right)}$ a band-pass transfer function centered on frequency $\mathit{f_k}$. $\mathbf{S_k}$ is a complex Hermitian matrix with real diagonal coefficients. The index $\mathit{k}$ in $\{1,2, \ldots, \mathit{P}\}$ is the frequency band index.

sPCA is the direct extension of classical PCA in the Fourier domain. In the fields of signal processing and systems control theory, this approach, which consists in identifying empirical linear dynamical relations between the variables of the system, is more commonly referred to as linear dynamical systems identification \cite{picci1986dynamic,georgiou2019dynamic}. While the essence of the approach always remains the same, it may be implemented in different ways. The various implementations of the method essentially differ in the way the cross-spectral matrix $\mathbf{S_k}$ is computed from the data and in the definition of the frequency bands $\{{\mathit{b_k}\left(f\right)}\}$.

As a counterpart of the greater flexibility (more degrees of freedom) of the sPCA method compared to the classical PCA, the sPCA is prone to overfitting when the cross-spectral coefficients are not robustly estimated. For example, from a set of $\mathit{L}$ observations (with $\mathit{L}$ finite) of $\mathit{N}$ variables, if no regularity is imposed on the cross-spectra, and if the number $\mathit{P}$ of independent frequency bands of the empirical cross-spectra is greater than or equal to $\mathit{L}$, one can always fit a "perfect" empirical dynamical relation (i.e. a transfer function) between any two variables $\mathbf{x_i}$ and $\mathbf{x_j}$ and thus obtain a rank-one cross-spectral matrix for each frequency band (see Appendix A). Therefore, there is a necessary trade-off between the spectral resolution (i.e. the number and width of frequency bands) and the robustness of the cross-spectral matrix estimation. The Bartlett and Welch periodogram methods \cite{bartlett1950periodogram,welch1967use,proakis2001digital} are the most frequently used methods for estimating robust (with low sampling variance) Fourier power spectra and cross-spectra from finite-length time series. They consist in splitting the time series into $\mathit{W}$ segments and then performing a Discrete Fourier Transform (DFT) for each segment. The Welch method uses overlapping segments while the Bartlett method uses non-overlapping segments. A robust estimation of the cross-spectral coefficients in each frequency band is obtained by averaging the complex cross-spectral coefficients obtained for each segment. As a tradeoff for the greater robustness of the computed cross-spectral coefficients because of reduced sample variance allowed by averaging, the Welch and Bartlett periodograms have reduced spectral resolutions since each one of the $\mathit{W}$ segments corresponds to a shorter time series (see Appendix A). 

While the Welch and Bartlett methods for computing periodograms are classified as non-parametric (as they do not rely on a parametric spectrum model), the user still needs to define the number of segments for the Bartlett method, plus the overlapping fraction for the Welch periodogram. The commonly used "modified" version of the Welsh periodogram also requires selecting a windowing function to be applied to each segment. All these methodological choices affect the computed cross-spectral matrix, in particular its rank and the decay rate of its eigenvalues (see Appendix A).

Other spectral estimation methods have been used in climate science to perform spectral PCAs. In \cite{mann1994global,mann1999oscillatory} Slepian tapers are used as weighting functions to compute cross-spectral quantities. The Slepian tapers are designed to minimize spectral leakage, i.e. compute cross-spectral coefficients over narrow frequency bands. The Slepian tapers method is adapted if, in the frequency band of interest, most of the variability of the system can be explained by a small number of PCs. Indeed, with this method, the rank of the empirical cross-spectral matrix is at most equal to the number of orthogonal tapers. In \cite{mann1994global,mann2020absence}, the method is implemented with $\mathrm{3}$ to $\mathrm{6}$ orthogonal tapers. In the cases when the system is too complex to be reduced to a few PCs, using the Slepian tapers method will inevitably lead to overfitting. The "multiwavelet" method \cite{lilly1995multiwavelet,park2000interannual} corresponds to a localized in time version of the Slepian taper method. The principal oscillation patterns (POPs) introduced by \cite{hasselmann1988pips} can be seen as a parametric version of sPCA where the PCs are autoregressive moving average (ARMA) processes.       

As a simple alternative to the spectral computation methods mentioned above, we propose to apply the sPCA by relying on the complex analytic Morlet wavelet to estimate the cross-spectral matrix. The continuous wavelet transform is obtained by convolving the analyzed signal with a basis of wavelet functions which are all dilated and translated versions of the same "mother" wavelet function. The Morlet mother wavelet \cite{morlet1982wave,addison2017illustrated} is defined as:
\begin{equation}
\Psi\left(t\right)=\pi^{-1/4}\left(e^{i2\pi f_0t}-e^{-{(2\pi  f_0)}^2}\right)e^{-t^2/2}
\end{equation}
which, if $2\pi\mathit{f_0}>5$ can be approximated as:
\begin{equation}
\Psi\left(t\right)=\pi^{-1/4}e^{i2\pi f_0t}e^{-t^2/2}
\end{equation}
The Morlet wavelet is therefore the complex exponential function $e^{i2\pi f_0t}=\cos{\left(2\pi f_0t\right)}+i\sin{\left(2\pi f_0t\right)}$ modulated by a Gaussian envelope. The continuous wavelet transform $\mathcal{W}_x(\mathit{\nu,t})$ of the signal $\mathit{x(t)}$ is defined as:
\begin{equation}
\mathcal{W}_x\left(\nu,t\right)=\frac{1}{\sqrt\nu}\int_{-\infty}^{+\infty}x\left(u\right)\Psi^\prime\left(\frac{u-t}{\nu}\right)du
\end{equation}
with $\mathit{\nu}$ the scale parameter in the wavelet time-scale domain. $\Psi^\prime(\mathit{u})$ designates the complex conjugate of $\Psi(u)$.

At scale  $\mathit{\nu_k}$, $\mathbf{W}_{\nu_\mathit{k}}$ is the $\mathit{N}$ by $\mathit{L}$ matrix of wavelet coefficients derived from the data matrix $\mathbf{X}$:
\begin{equation}
\boldsymbol{W_{\nu_\mathit{k}}}=
\left[\begin{matrix}
      \mathcal{W}_{x_1}\left(\nu_k,t_1\right)&\cdots&
      \mathcal{W}_{x_1}\left(\nu_k,t_L\right)\\\vdots&\ddots&\vdots\\
      \mathcal{W}_{x_N}\left(\nu_k,t_1\right)&\cdots&
      \mathcal{W}_{x_N}\left(\nu_k,t_L\right)\\
      \end{matrix}\right]
\end{equation}
Each scale $\mathit{\nu_k}$ in the wavelet scale domain corresponds to a frequency band $\mathit{b_k}$ in the Fourier frequency domain with the central frequency of $\mathit{b_k}$ being $\mathit{f_k}=\frac{f_0}{\mathit{\nu_k}}$ (see Appendix B). Because of the correspondence between the Morlet wavelet transform and the Fourier transform (equations 7, 8 and Appendix B), the empirical sample cross-spectral matrix for the frequency band $\mathit{b_k}$ can be computed as:
\begin{equation}
\overline{\mathbf{S_k}}=\frac{1}{N}\mathbf{W_{\nu_\mathit{k}}}\mathbf{W_{\nu_\mathit{k}}}^\prime
\end{equation}
The Morlet wavelet method to compute the cross-spectral matrix can be seen as similar to using a modified Welch periodogram with a Gaussian window whose length varies inversely proportional to the frequency. Our proposed method can also be related to the “multiwavelet” method \cite{lilly1995multiwavelet,park2000interannual}, except that in our case, robustness is obtained through temporal integration rather than using multiple wavelets with mutually exclusive frequency support. 
Similarly to the classical real-value PCA case, the complex eigenvectors $\mathbf{u_{i,k}}$ of the $\overline{\mathbf{S_k}}$ matrix are the solutions of the equation:
\begin{equation}
\overline{\mathbf{S_k}}\mathbf{u_{i,k}}={\lambda_{i,k}}^2\ \mathbf{u_{i,k}}
\end{equation}
The eigenvalues ${\lambda_{i,k}}^2$ of $\overline{\mathbf{S_k}}$ are real from the fact that $\overline{\mathbf{S_k}}$ is by construction a Hermitian matrix. The PC series of wavelet coefficients associated with the eigenvector $\mathbf{u_{i,k}}$ at the scale $\mathit{\nu_k}$ is:
\begin{equation}
\boldsymbol{\kappa}_\mathbf{i,k}=\mathbf{W}_{\nu_k}\mathbf{u_{i, k}}
\end{equation}
We can then reconstruct the time series corresponding to the $\mathit{i}^{th}$ wavelet PC at scale $\mathit{\nu_k}$ through an inverse wavelet transform. We can also combine several wavelet PCs at various scales and then apply the inverse wavelet transform to reconstruct a time series which encapsulates the variability of the original signal within any desired frequency band (range of scales). However, one shall note that, while the phase shift between two time series in a given frequency band is physically interpretable, the absolute value of $\mathrm{arg\left(\mathbf{\kappa_{i,k}}\right)}$ is arbitrary and not meaningful (indeed, if $\mathbf{u_{i,k}}$ is a solution of equation (11), so is $e^{i\theta}\mathbf{u_{i,k}}$ for any value of $\theta$). Therefore, to make the reconstructed time series interpretable when combining several wavelet PCs, we shift the phase of each wavelet PC such that it is aligned in phase with the highest contributing series of wavelet coefficient (i.e. the variable having the highest absolute weight in the corresponding eigenvector) before performing the inverse wavelet transform. One shall also note that the inverse wavelet transform with the Morlet wavelet is not an exact reconstruction; however, in practice we find that the time series are well enough reconstructed to be interpretable. 

The only parameter which has to be selected by the user for the implementation of the Morlet wavelet sPCA is the central frequency $\mathit{f}_0$ of the mother wavelet. It is generally chosen between $\mathrm{0.8}$ and $\mathrm{1}$; the value $\mathit{f}_0=0.849$ $(=\sqrt{1/2\ln{\left(2\right)}})$ can be chosen such as the magnitude of the second highest peak of the wavelet is half the magnitude of the highest peak (central peak). In practice, values between $\mathrm{0.8}$ and $\mathrm{1}$ will produce similar spectra. Taking a higher than $\mathrm{1}$ $\mathit{f}_0$ value will lead to a narrower spectral bandwidth (higher frequency resolution). It will also lead to poorer localization of the wavelet coefficients in the time domain (lower time resolution) which in practice translates into wavelet coefficients more correlated in time, i.e. fewer independent samples for estimating the cross-spectral coefficients, meaning less robust empirical cross-spectra and higher risk of overfitting. As for the periodogram methods, the tradeoff between frequency resolution and time resolution / number of independent samples is related to the Heisenberg–Gabor limit \cite{gabor1946theory,addison2017illustrated}. One of the interesting properties of the Morlet wavelet is that it actually reaches the Heisenberg–Gabor limit, therefore allowing the best possible time resolution for a given frequency resolution. Because the Gaussian envelope of the Morlet wavelet shrinks at higher frequencies (finer scales), the cross-spectral coefficients are more robustly estimated at higher frequencies, allowing to potentially extract robust coherent high-frequency modes even if their amplitude is low relatively to the noise, the cost of this being reduced frequency resolution at higher frequencies. One shall note that the wavelet coefficients are potentially affected by edge effects; these can be avoided by considering only coefficients outside of the cone of influence or by performing appropriate padding of the time series before the wavelet transform (see Appendix C).
%
%
%
%
%
%
\subsection{Rotation of eigenvectors with phase regularization for physical interpretability}

In the case when the PCA identifies two or more PCs of equal importance (i.e. associated with equal eigenvalues $\mathit{\lambda_i)}$ in a given frequency band, any orthonormal basis of the subspace generated by these orthogonal PCs is a valid solution to the SVD problem. Therefore, one can arbitrarily select any rotated basis of the generated subspace to project the data. In the case when the eigenvalues are not identical but close to each other, a very small alteration of the dataset (e.g., noise in the data) may lead to very different empirical PCs. While the information retained by the PC reduction would remain the same under any rotation of the basis formed by the PCs, this complicates the interpretation and physical attribution of the PCs. This fact had already been pointed out by \cite{wallace1971spectral} and \cite{wallace1972aempirical}, among others, as a limitation of the sPCA technique. It can also affect classical PCA (see e.g. the “rule of thumb” in \cite{north1982sampling}) and is one of the reasons why rotated PCA methods have been proposed \cite{richman1986rotation}. Rotated PCA methods aim at finding a rotation of the eigenvectors (or rotation of the PCs) maximizing a given criterion (or minimizing a cost function). The varimax method \cite{kaiser1958varimax} is for example the most widely used form of rotated PCA. It consists in finding the rotation which maximizes the sum of the squared correlations between the original variables and the rotated PCs. While rotated PCAs have often been used in climate science \cite{mestas1999rotated, lian2012evaluation,chen2017pairwise}, they have rarely been applied in a complex domain such as the Fourier domain, and, in these rare examples \cite{wallace1972aempirical,bloomfield1994orthogonal,bueso2020nonlinear}, only with the varimax or promax criteria, which ignore the phase information. 

When each variable of the system can be attributed to a spatial location, the spatial structure of the eigenvectors can be analyzed and used as a criterion to find interpretable and meaningful rotated PCs. For the sPCA, the spatial structure of the phase of the complex eigenvectors is particularly informative, especially when one seeks to identify dynamical modes with propagation effect. Indeed, for a propagating wave, the phase is expected to be a linear (or at least locally linear) function of space. We therefore propose to use a metric characterizing the spatial regularity of the phase of the rotated eigenvectors as a criterion to define physically interpretable rotated PCs from the sPCA.
Let us consider $\mathbf{u_1}$ and $\mathbf{u_2}$ two unit-norm eigenvectors of the cross-spectral matrix $\mathbf{S_k}$ associated with the $\lambda_1$ and $\lambda_2$ eigenvalues with $\lambda_1\approx\lambda_2$. In the subspace generated by $\mathbf{u_1}$ and $\mathbf{u_2}$, any unit-norm vector, can be written as:
\begin{equation}
\mathbf{u}_{\theta,\varphi,\omega}=e^{i\omega}(cos\left(\theta\right)\mathbf{u}_\mathbf{1}+sin\left(\theta\right)e^{i\varphi}\mathbf{u}_\mathbf{2})
\end{equation}
\begin{figure}[!t]
\centering
\includegraphics[width=1\textwidth]{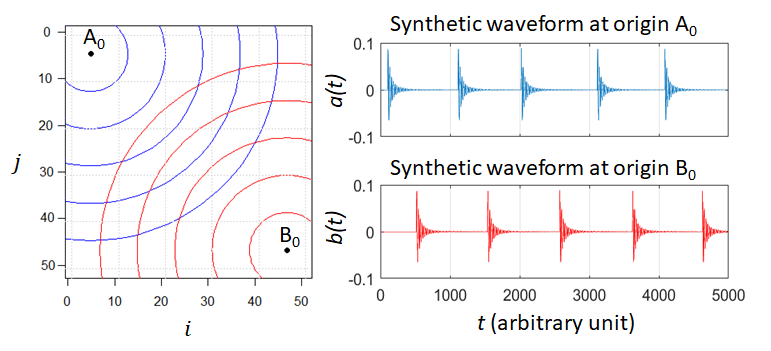}
\caption{Wave propagation example. (Left) Representation of the simulated $\mathrm{51}$ by $\mathrm{51}$ domain with the origins $\mathrm{A_0}$ and $\mathrm{B_0}$ of the two waves $\mathit{a(t)}$ and $\mathit{b(t)}$ marked; (right) synthetic waveforms $\mathit{a(t)}$ and $\mathit{b(t)}$.}
\label{fig1}
\end{figure}
The vector $\mathbf{u}_\mathit{\theta,\varphi,\omega}$ is a complex vector belonging to $\mathbb{C}^N$. All the vectors of the form defined in equation (13) are rotations of each other in the subspace generated by $\mathbf{u_1}$ and $\mathbf{u_2}$. We noted previously that the absolute value of the argument of an eigenvector is arbitrary, this is also true for the rotated eigenvectors, therefore we can arbitrarily impose $\omega=0$ and consider only the vectors of the form:
\begin{equation}
\mathbf{u}_{\theta,\varphi}=cos\left(\theta\right)\mathbf{u}_\mathbf{1}+sin\left(\theta\right)e^{i\varphi}\mathbf{u}_\mathbf{2}
\end{equation}
In the case when all $\mathit{N}$ time series forming the $\mathbf{X}$ dataset are associated to a spatial location (in a $\mathit{D}$-dimensional space), $\mathbf{u}_\mathit{\theta,\varphi}$ can be represented as a complex function of space $\mathbf{u}_\mathit{\theta,\varphi}\left(\mathbf{y}\right)$; $\mathbf{y}$ being a $\mathit{D}$-dimensional coordinate vector. We can then define:
\begin{equation}
J\left(\mathbf{u}_{\theta,\varphi}\right)=\int_{\mathbb{R}^D}^{\ }{|\mathrm{\Delta}\arg(\mathbf{u}_{\theta,\varphi}\left(\mathbf{y}\right))|\mathrm{d}\mathbf{y}}
\end{equation}
where $\mathrm{\Delta}$ denotes the Laplacian operator in $\mathit{D}$ dimensions. If $\arg(\mathbf{u}_{\mathit{\theta},\mathit{\varphi}}\left(\mathbf{y}\right))$ is a linear function of space, then $\mathrm{\Delta}\arg(\mathbf{u}_{\theta,\varphi}\left(\mathbf{y}\right))$ is null everywhere and $J(\mathbf{u}_{\theta,\varphi})=0$. Therefore, to identify rotated PCs corresponding to linearly propagating waves, we look for the rotated vector:
\begin{equation}
\mathbf{u_{r1}}=\argmin{\mathbf{u}_{\theta,\varphi}}(J(\mathbf{u}_{\theta,\varphi})) 
\end{equation}
We note that the minimization of $J\left(\mathbf{u}_{\theta,\varphi}\right)$ is also meaningful in the case of stationary waves for which $\mathrm{arg}(\mathbf{u}_{\mathbf{r1}}\left(\mathit{y}\right))$ is expected to be constant. In practice, with a finite number $\mathit{N}$ of time series corresponding to discrete locations, $J_{\theta,\varphi}$ is approximated as:
\begin{equation}
\hat{J}(\mathbf{u}_{\theta,\varphi})=
\sum_{i=1}^{N}
{|\hat{\mathrm{\Delta}}\arg(\mathbf{u}_{\theta,\varphi}\left(\mathbf{y}_\mathbf{i}\right))|}
\end{equation}
where $\hat{\mathrm{\Delta}}$ is a discrete Laplacian operator (see Appendix D).  

%
%
%
%
%
%
\section{Demonstration of the rotated spectral PCA (rsPCA) in a synthetic example}

We demonstrate the possibilities of the proposed methodology by applying it to a synthetic data generated by a simple numerical model simulating two waves propagating in opposite directions on a $\mathrm{2D}$ plane. The magnitudes, phase and propagation speed of the waves are controlled, as is the noise of the system. The simulated system corresponds to $\mathit{N}=\mathrm{2601}$ time series of length $\mathit{L}=\mathrm{5000}$ samples (with arbitrary time unit), spatially distributed on a $\mathrm{51}$ by $\mathrm{51}$ spatial grid (Figure 1). At the $\mathit{(i, j)}$ location (with $\mathit{i}$ and $\mathit{j}$ in ${\{1, 2, \ldots, 51}\}$), the time series $\mathit{x_{i,j}(t)}$ is:
\begin{equation}
\begin{split}
x_{i,j} \left(t\right) = & 
\frac{1}{d_0 + d_a}\ a\left(t + \frac{d_a}{c}\right)
 + \\ & \frac{1}{d_0 + d_b}\gamma_1\ b\left(t + \frac{d_b}{c}\right)
 + \gamma_2N_{i,j}\left(t\right)
\end{split}
\end{equation}
with:
\begin{equation}
d_a = \sqrt{{(i - 5)}^2 + {(j - 5)}^2}
\end{equation}
\begin{equation}
d_b = \sqrt{{(i - 47)}^2 + {(j - 47)}^2}
\end{equation}
\begin{figure}[!b]
\centering
\includegraphics[width=1\textwidth]{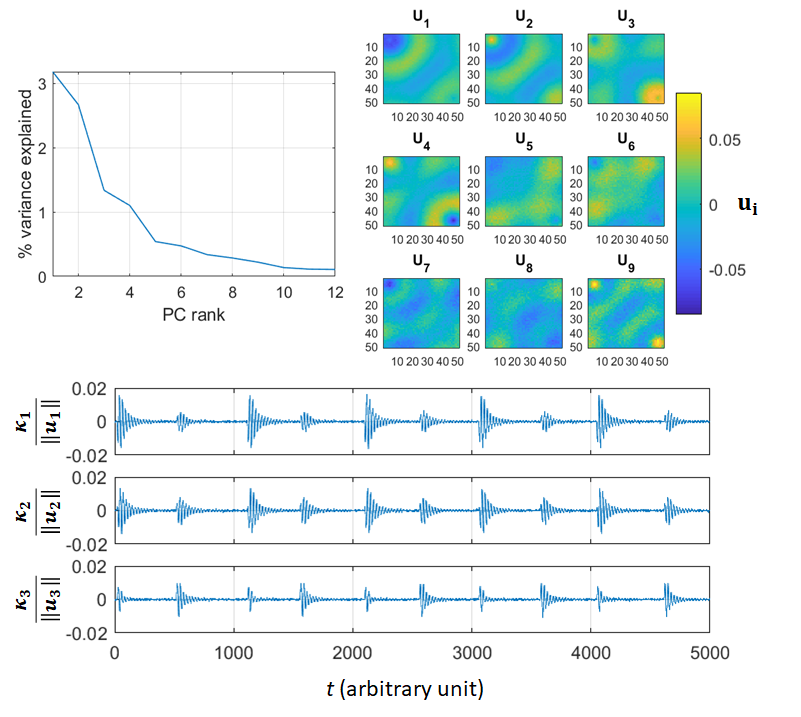}
\caption{Classical PCA applied to the wave propagation example. (Top left) Fraction of the variance of the synthetic system explained by the first $\mathrm{12}$ PCs for the classical PCA. (Top right) Spatial representation of the eigenvectors associated with the first $\mathrm{9}$ PCs. (Bottom) First $\mathrm{3}$ PC time series. For this setup, $\gamma_1\mathrm{=0.75}$; the waves $\mathit{a(t)}$ and $\mathit{b(t)}$ and the noise account respectively for $\mathrm{6\%}$, $\mathrm{4\%}$ and $\mathrm{90\%}$ of the variance of the system.}
\label{fig2}
\end{figure} 
The functions $\mathit{a(t)}$ and $\mathit{b(t)}$, shown on Figure 1, are two intermittent waveforms. They oscillate at the same frequency, but their relative phase is not coherent. The $\mathit{a(t)}$ signal originates from the $\mathrm{(5, 5)}$ grid point (point $\mathrm{A_0}$) and propagates linearly through the domain. Similarly, the $\mathit{b(t)}$ signal originates from the $\mathrm{(47, 47)}$ grid point (point $\mathrm{B_0}$) and propagates linearly through the domain. The parameter $\mathit{c}$ controls the propagation speed of the $\mathit{a(t)}$ and $\mathit{b(t)}$ signals. The parameters $\gamma_1$ and $\gamma_2$ control the relative amplitude of the two waves and of the noise. Two setups are presented below, 1) the two waves having different amplitudes ($\gamma_1=0.75), 2$) the two waves having the same amplitude ($\gamma_1=1$). For both setups, $\mathit{c}=\mathrm{1.2}$.

In the first setup, $\mathit{\gamma_1}$ is set to $\mathrm{0.75}$ and $\mathit{\gamma_2}$ is set to 0.11, such that the $\mathit{a(t)}$ and $\mathit{b(t)}$ signals and the noise $\mathit{N_{i,j}\left(t\right)}$ account respectively for  $\mathrm{6\%}$, $\mathrm{4\%}$ and $\mathrm{90\%}$ of the variance of the system. The classical PCA is applied to identify the principal modes of variation of the $\mathrm{2601}$ time series of the synthetic system. The result of the PCA (fraction of variance explained by the first $\mathrm{12}$ PCs, spatial structure of the first $\mathrm{9}$ eigenvectors and first $\mathrm{3}$ PC time series) is shown on Figure 2. One will note that no less than $\mathrm{9}$ PCs are needed to capture $\mathrm{10\%}$ of the variance (i.e. the fraction of the variance explained by the two propagating waves $\mathit{a(t)}$ and $\mathit{b(t)}$), and that none of the first $\mathrm{9}$ eigenvectors can be related specifically to one of the two waves but all $\mathrm{9}$ are rather a combination of both $\mathit{a(t)}$ and $\mathit{b(t)}$. This illustrates the poor ability of the classical PCA to capture dynamical linear relations between the variables when propagation effects are involved. The same system is then analyzed with the wavelet-based sPCA method. 
\begin{figure}[!t]
\centering
\includegraphics[width=1\textwidth]{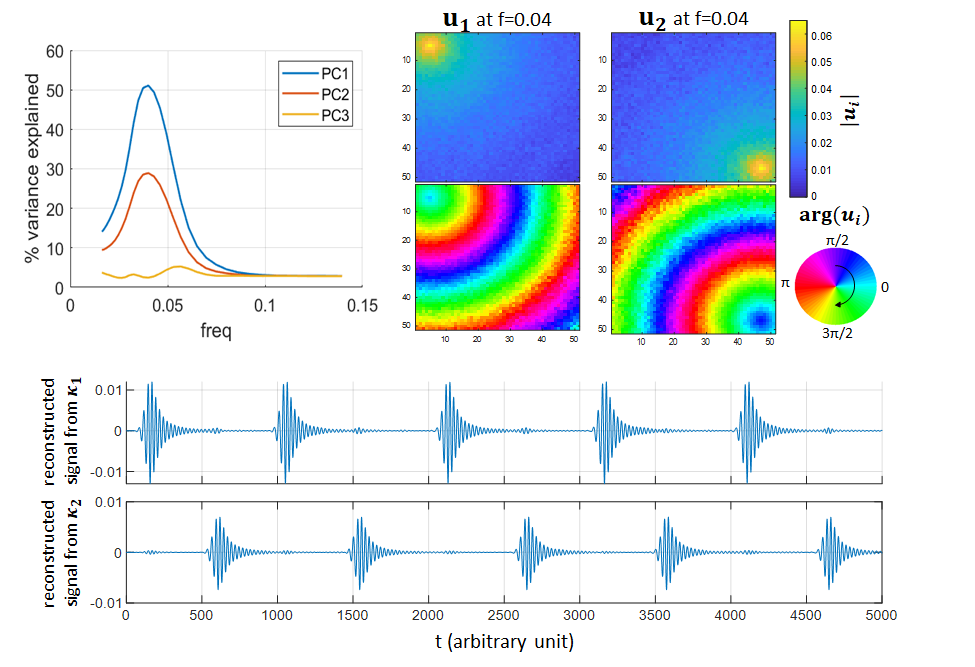}
\caption{Spectral PCA applied to the wave propagation example. (Top left) Fraction of the variance of the synthetic system explained by the first $\mathrm{3}$ PCs of the sPCA as a function of the frequency. (Top right) Spatial representation of the eigenvectors (phase and modulus) associated with the first $\mathrm{2}$ PCs at frequency $\mathit{f_k}\mathrm{=0.04}$. The circular arrow in the colorscale of the phase indicates the direction of propagation of the extracted waves. (Bottom) Reconstructed signal from the combination of the first (/second) wavelet PCs at all frequencies between $\mathrm{0.026}$ and $\mathrm{0.053}$; wavelet PCs are phase adjusted before reconstruction. For this setup, $\gamma_1\mathrm{=0.75}$; the waves $\mathit{a(t)}$ and $\mathit{b(t)}$ and the noise account respectively for $\mathrm{6\%}$, $\mathrm{4\%}$ and $\mathrm{90\%}$ of the variance of the system.}
\label{fig3}
\end{figure} 

The results are shown on Figure 3. For the frequency band centered at the frequency $\mathit{f_k}=\mathrm{0.04}$ the first two PCs explain respectively $\mathrm{46\%}$ and $\mathrm{26\%}$ of the system’s variance. The phase and modulus structure of the two corresponding eigenvectors accurately describe the propagation of $\mathit{a(t)}$ and $\mathit{b(t)}$ with good separation of the two waves. The signal reconstructed through inverse wavelet transform from the first wavelet PCs at frequencies $\mathrm{0.026}$ to $\mathrm{0.053}$ corresponds to the $\mathit{a(t)}$ signal while the signal reconstructed from the second PCs within the same frequency range corresponds to the $\mathit{b(t)}$ signal. We can estimate the propagation speed from the phase information. Considering that a phase shift of $\mathrm{2\pi}$ at $\mathit{f_k}=\mathrm{0.04}$ corresponds to $\mathrm{30}$ grid increments we estimate $\hat{c}=30\times0.04=1.2$, which is an accurate estimation of the true propagation speed. The clockwise progression of the phase corresponds to a forward delay in time.  
\begin{figure}[!t]
\centering
\includegraphics[width=1\textwidth]{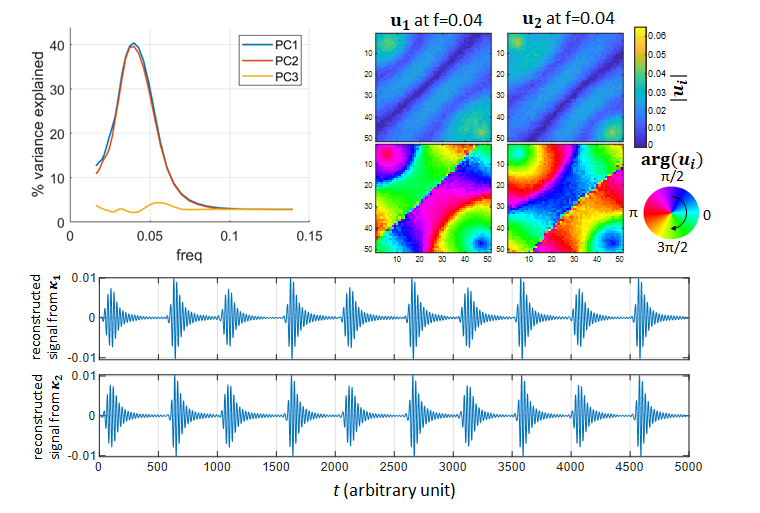}
\caption{Same as Fig. 3 with $\gamma_1\mathrm{=1}$. In this setup the waves $\mathit{a(t)}$ and $\mathit{b(t)}$ and the noise account respectively for $\mathrm{5\%}$, $\mathrm{5\%}$ and $\mathrm{90\%}$ of the variance of the system.}
\label{fig4}
\end{figure} 
\begin{figure*}[!hbtp]
\centering
\includegraphics{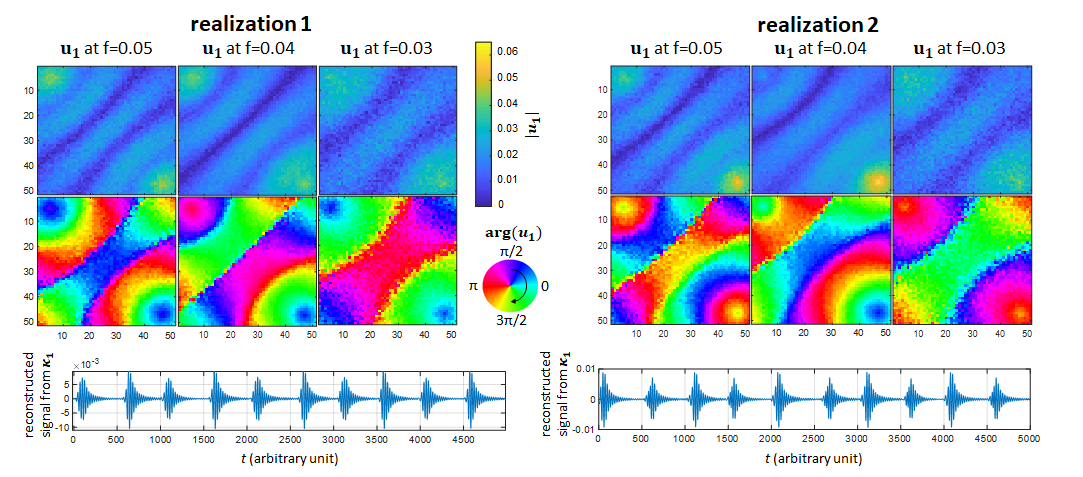}
\caption{ Instability of the sPCA solution across frequencies and sensitivity to noise when several modes of similar amplitude coexist within the same frequency band. (Top) First eigenvector $\mathbf{u_1}$ (modulus and phase) identified by the sPCA at frequencies 0.03, 0.04 and 0.05 with two different realizations of the random noise $\mathit{N_{i,j}}\left(t\right)$. The circular arrows in the colorscales of the phase indicate the direction of propagation of the extracted waves. (Bottom) Reconstructed signals from the combination of the first wavelet PCs at all frequencies between $\mathrm{0.026}$ and $\mathrm{0.053}$. For the two realizations,$\gamma_1\mathrm{=1}$; the waves $\mathit{a(t)}$ and $\mathit{b(t)}$ and the noise account respectively for $\mathrm{5\%}$, $\mathrm{5\%}$ and $\mathrm{90\%}$ of the variance of the system.}
\label{fig5}
\end{figure*}
In the second setup, $\gamma_1$ is set to $\mathrm{1}$, such that both $\mathit{a}$ and $\mathit{b}$ waves have same total energy, and $\gamma_2$ is adjusted to $\mathrm{0.13}$ such that the noise still accounts for $\mathrm{90\%}$ of the variance of the system. The results of the sPCA for this setup are shown on Figure 4. In this case, sPCA is able to determine that two modes of equal importance, explaining together $\mathrm{74\%}$ of the system’s variance, exist around the frequency $\mathrm{0.04}$. Figure 5 shows the first eigenvector $\mathbf{u_1}$ identified by the sPCA at frequencies $\mathrm{0.03}$, $\mathrm{0.04}$ and $\mathrm{0.05}$ for the same synthetic system as previously, with two different realizations of the random noise $\mathit{N_{i,j}\left(t\right)}$. One can see that the different realizations of the noise produce different eigenvectors $\mathbf{u_1}$ and also that the spatial structure of the eigenvectors may vary significantly across nearby frequency bands. Additionally to being extremely sensitive to noise and frequency variations, the eigenvectors $\mathbf{u_1}$ and $\mathbf{u_2}$ are not easily interpretable since both of them are always reflecting a complex combination of the waves $\mathit{a}$ and $\mathit{b}$. One should however consider the fact that the subspace generated by the first two PCs is the same for all these cases, and, together, the first two PCs always capture the signal corresponding to the superposition of the waves $\mathit{a}$ and $\mathit{b}$.

We then perform a rotation of the eigenvectors in the subspace generated by $\mathbf{u_1}$ and $\mathbf{u_2}$ at frequency $\mathrm{0.04}$ as described in section 2.4 to find the rotated vector $\mathbf{u_{r1}}$ for which $\hat{J}(\mathbf{u}_{\theta,\varphi})$ is minimal. Figure 6 shows the value of the criterion $\hat{J}(\mathbf{u}_{\theta,\varphi})$ as a function of the rotation parameters $\mathrm{\theta}$ and $\mathrm{\varphi}$, as well as the rotated vector $\mathbf{u_{r1}}$ that minimizes $\hat{J}(\mathbf{u}_{\theta,\varphi})$ and the $\mathbf{u_{r2}}$ orthogonal to $\mathbf{u_{r1}}$ in the plane defined by $\mathbf{u_1}$ and $\mathbf{u_2}$. One can see that the rotated vector $\mathbf{u_{r1}}$ corresponds to the wave $\mathit{a(t)}$ while $\mathbf{u_{r2}}$ corresponds to the wave $\mathit{b(t)}$. One can also note that $\hat{J}(\mathbf{u}_{\theta,\varphi})$ shows two local minima in the $\{\mathrm{\theta,\varphi}\}$ parameter space. The second local minimum corresponds to the symmetric solution (not shown), i.e. $\mathbf{u_{r1}}$ associated to the wave $\mathit{b(t)}$ and $\mathbf{u_{r2}}$ associated to the wave $\mathit{a(t)}$. 
%
%
%
%
%
%
\begin{figure}[!hbtp]
\centering
\includegraphics[width=1\textwidth]{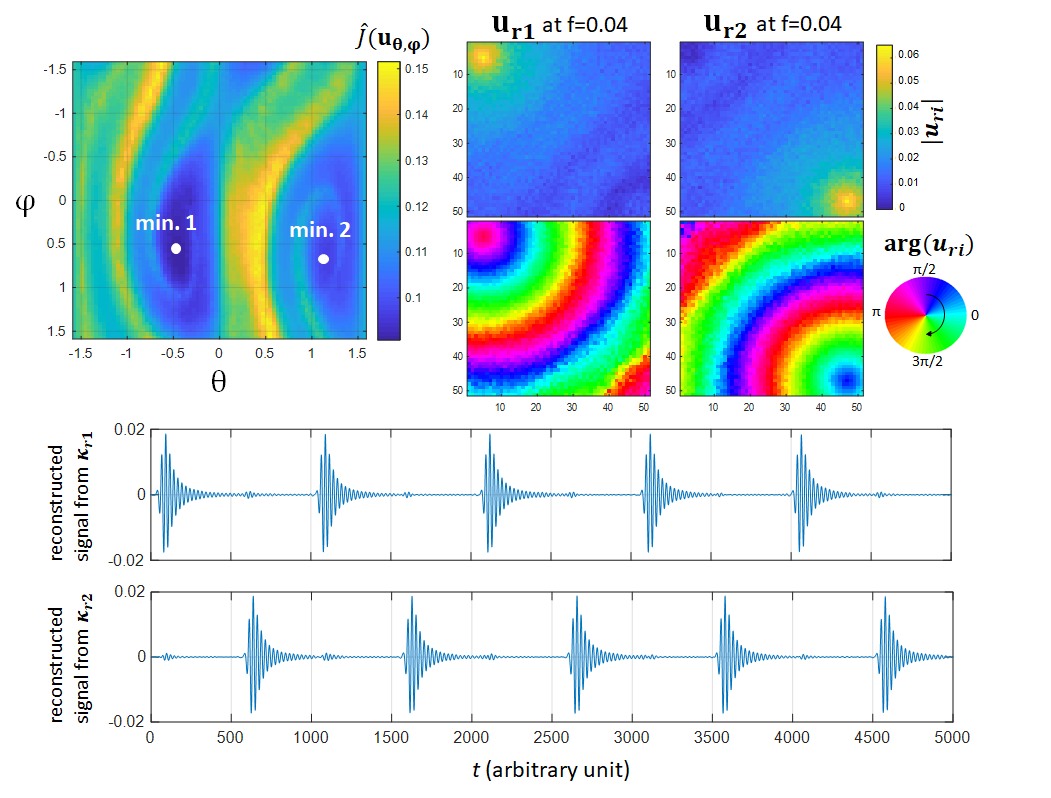}
\caption{Rotated spectral PCA (rsPCA) applied to the wave propagation example. (Top left) Criterion $\hat{J}(\mathbf{u}_{\theta,\varphi})$ as a function of the rotation parameters $\mathrm{\vartheta}$ and $\mathrm{\varphi}$ at frequency $\mathit{f_k}\mathrm{=0.04}$. (Top right) Rotated vector $\mathbf{u_{r1}}=\argmin{\mathbf{u}_{\theta,\varphi}}(\hat{J}(\mathbf{u}_{\theta,\varphi}))$ and $\mathbf{u_{r2}}$ orthogonal to $\mathbf{u_{r1}}$ at frequency $\mathit{f_k}\mathrm{=0.04}$ (minimum $\mathrm{1}$ on the left panel). The circular arrow in the colorscale of the phase indicates the direction of propagation of the extracted waves. (Bottom) Reconstructed signal from the combination of the first (/second) wavelet PC at all frequencies between  $\mathrm{0.026}$ and $\mathrm{0.053}$. It is noted that the second local minimum $\mathrm{2}$ on the top left panel corresponds to the symmetric solution (not shown), i.e., $\mathbf{u_{r1}}$ associated to the wave $\mathit{b(t)}$ and $\mathbf{u_{r2}}$ associated to the wave $\mathit{a(t)}$.}
\label{fig6}
\end{figure} 
\section{Analysis of sea surface temperature over the Pacific Ocean}
In this section, we apply the rsPCA methodology to monthly SST series over the Pacific Ocean, with the aim of identifying the principal modes of SST variability across a range of scales/frequencies. Although much progress has been made over the last decades in identifying and physically interpreting the dominant SST modes of the Pacific, such as the Pacific Decadal Oscillation (PDO) \cite{mantua1997pacific,mantua2002pacific,newman2016pacific}, and the El Ni\~{n}o-Southern Oscillation (ENSO)\cite{trenberth1997definition,wang2017nino}, important open questions still remain. Specifically, SST modes which correspond to finer timescales (seasonal to sub-seasonal, S2S) have received less attention, since they are usually masked by decadal modes in classical PCA studies because they correspond to lower variance of the system, or because, as dynamical modes with non-negligible propagation effects, they are not well identified by the classical PCA. Better tracking, understanding, and modelling of S2S patterns of SST variability can improve prediction skill of regional hydroclimate, an exigent problem for many regions around the globe \cite{board2016next}. Moreover, the longer and higher quality observational records available today allow for more reliable investigation of the changing SST Pacific dynamics towards unraveling possible new emerging modes of forced climate variability \cite{yeh2009nino,fan2018climate,mamalakis2018new,mamalakis2019reply}. 

At sub-annual scales, the leading principal modes correspond to almost equal variance (not the case for longer timescales or low frequency modes, as will be shown below), which may complicate the interpretation and physical attribution of these modes, and make results sensitive to measurement uncertainties. As such, studying sub-annual SST dynamics presents a demanding setting in which to test the efficiency of the proposed rsPCA methodology in separating coherent patterns of variability when the underlying PCs are of equal importance (i.e., there is not a single dominant mode) and when robust extraction of propagating mode dynamics is needed for interpretation.   
%
%
%
%
%
%
\subsection{SST data used for analysis}
The analyzed data consist of a record of $\mathrm{79}$ years $\mathrm{(1940-2018)}$ of monthly SST ($\mathrm{948}$ time steps) over the Pacific Ocean from the second version of the COBE SST dataset (Centennial in situ Observation-Based Estimates of SST, \cite{hirahara2014centennial}). The original data is provided on $\mathrm{1\degree}$ by $\mathrm{1\degree}$ latitude / longitude grid. For the purpose of this study, the SST data from the COBE dataset is re-projected on $\mathrm{220}$ km by $\mathrm{220}$ km equal-area pixels ($\mathrm{3960}$ pixels in total) through a Mollweide projection \cite{snyder1977comparison}. The re-projection must be performed before applying the PCA, otherwise, high-latitude areas are given excessive weight in terms of contribution to the system’s variance. Because the temporal variations of the SST are dominated by the seasonal cycle, which is not what we are trying to characterize here, we remove the seasonal variations of the SST by subtracting the climatic monthly mean (mean over $\mathrm{79}$ years) and normalizing by the monthly standard deviation in each pixel. We thus obtain a unit-variance zero-mean time series of SST monthly anomalies.
\begin{figure}[!b]
\centering
\includegraphics[width=1\textwidth]{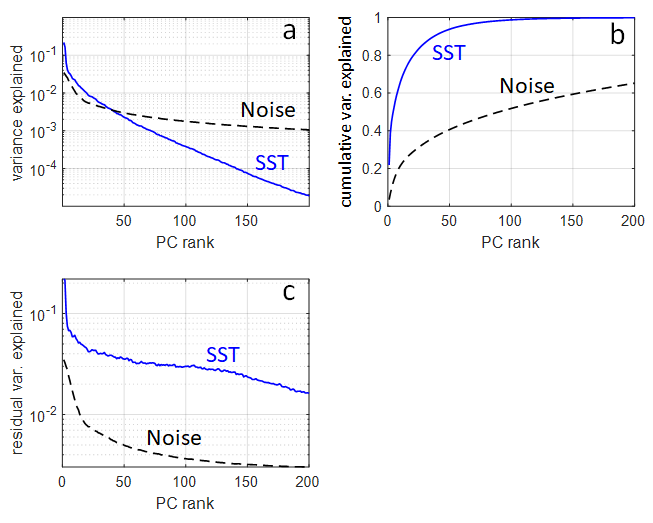}
\caption{Classical PCA applied to SST monthly anomaly series over the Pacific Ocean. (a, solid blue curve) Fraction of the system variance explained by the first $\mathrm{150}$ PCs, (b, solid blue curve) cumulative variance explained as a function of the PC rank, (c, solid blue curve) fraction of the system’s residual variance explained by the first $\mathrm{150}$ PCs (the residual variance at rank $\mathit{n}$ is the variance of the system minus the variance explained by all the PCs of rank lower than $\mathit{n}$). In panels a, b and c, the dashed black curves show what would be obtained if all $\mathrm{3960}$ time series were independent random time series (colored noise with a Fourier power spectrum identical to the power spectrum of the SST anomalies time series).}
\label{fig7}
\end{figure} 
The studied system is therefore composed of $\mathrm{3960}$ time series of SST anomalies of length $\mathrm{948}$ months (data matrix $\mathbf{X}$ with dimension $\mathrm{3960}$ by $\mathrm{948}$). One must note that the number of time series (i.e. the number of degrees of freedom of the PCA) is higher than the length of the time series (i.e. the number of constraints). Because the empirical rank of the system cannot exceed the number of constraints, the rank of the covariance and cross-spectral matrices cannot exceed $\mathrm{948}$ in our case (see Appendix A).  
%
%
%
%
%
%
\subsection{Results}
Similarly to what can be found in several published studies \cite{weare1976empirical,messie2011global}, Figures 7 and 8 show the results obtained by applying the classical PCA to the Pacific Ocean SST anomalies. From Figure 7, we note that a large number of PCs would be necessary to comprehensively describe the complexity of the Pacific Ocean's SST (for example, more than $\mathrm{20}$ PCs are needed to explain $\mathrm{80\%}$ of the variability). We also note that each one of the first $\mathrm{150}$ PCs explains more residual variance than would be expected for a set of spatially uncorrelated time series (noise), which is not surprising considering that the SST anomalies are obviously spatially correlated. The PC series and corresponding eigenvectors are shown in Figure 8. The first two PCs, accounting respectively for $\mathrm{22\%}$ and $\mathrm{17\%}$ of the system’s variability, are related to the ENSO and to anthropogenic climate change. The third and fourth PCs correspond to $\mathrm{6\%}$ and $\mathrm{4\%}$ of the variance, respectively, and their signal is stronger mostly over the northern Pacific. These patterns seem to resemble the patterns of the North Pacific gyre oscillation \cite{di2008north,di2009nutrient} and the PDO (see the decadal variability of the fourth PC series). All these physical modes of variability and change are not perfectly separated by the PCA, as the first two PCs both reflect part of the ENSO and climate change signals, and the third and fourth PCs reflect signals of the PDO and the North Pacific gyre oscillation. In this case, a rotation of the PCs with an adapted criterion may allow a better physical attribution and interpretation. We also note that all first four modes identified by the classical PCA are low-frequency modes corresponding to inter-annual to decadal time scales.

\begin{figure}[!t]
\centering
\includegraphics[width=1\textwidth]{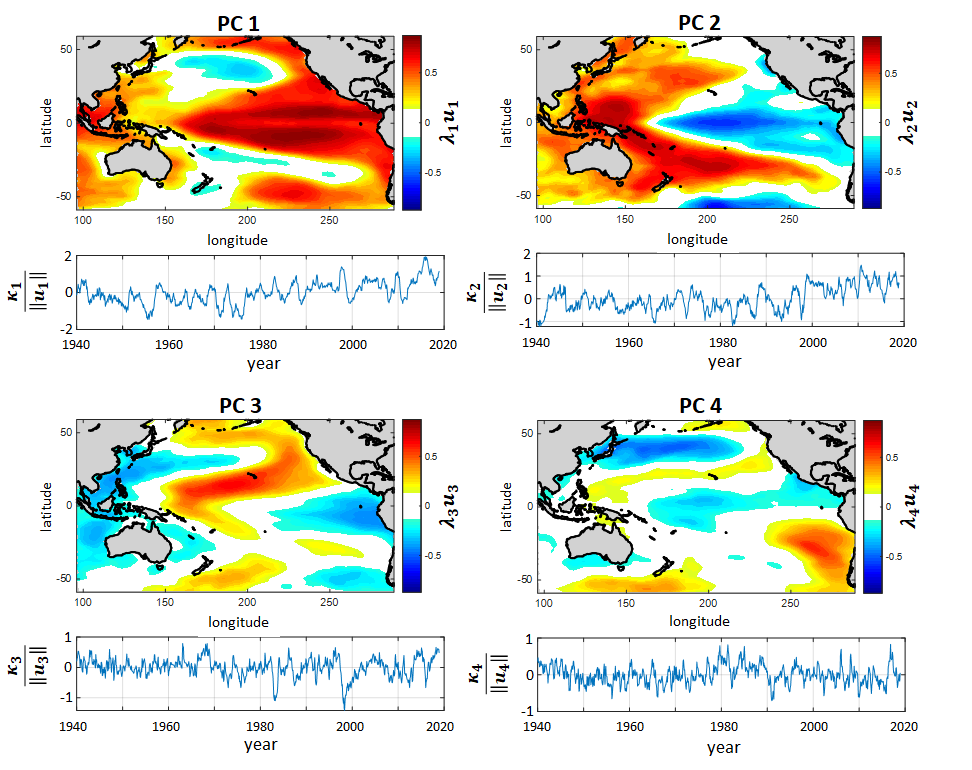}
\caption{First 4 PCs $\mathrm{\boldsymbol{\kappa_i}}$ and associated eigenvectors $\mathbf{u_i}$ resulting from the classical PCA applied to the SST monthly anomalies time series. The first four PCs explain respectively $\mathrm{22\%}$, $\mathrm{17\%}$, $\mathrm{6\%}$ and $\mathrm{4\%}$ of the system's variability.}
\label{fig8}
\end{figure} 

In order to unmix modes originating from different frequency bands and also explore the SST variability on sub-annual scales, we then apply the wavelet-based sPCA to the SST data. Figure 9 shows how much of the variance is explained by the first four spectral PCs as a function of the frequency $\mathit{f_k}$. For frequencies lower than $\mathrm{0.5}$ cycle per year, the corresponding first PCs explain more than $\mathrm{25\%}$ of the system variability. The fraction of variance explained by the first PC peaks above $\mathrm{50\%}$ at frequencies $\mathrm{1/3.4}$, $\mathrm{1/5.5}$, and $\mathrm{1/11.8}\ \mathrm{yr}^{-1}$. These results show that the Pacific SST variability at lower frequencies can be efficiently characterized by a small number of principal modes.
\begin{figure}[!b]
\centering
\includegraphics[width=1\textwidth]{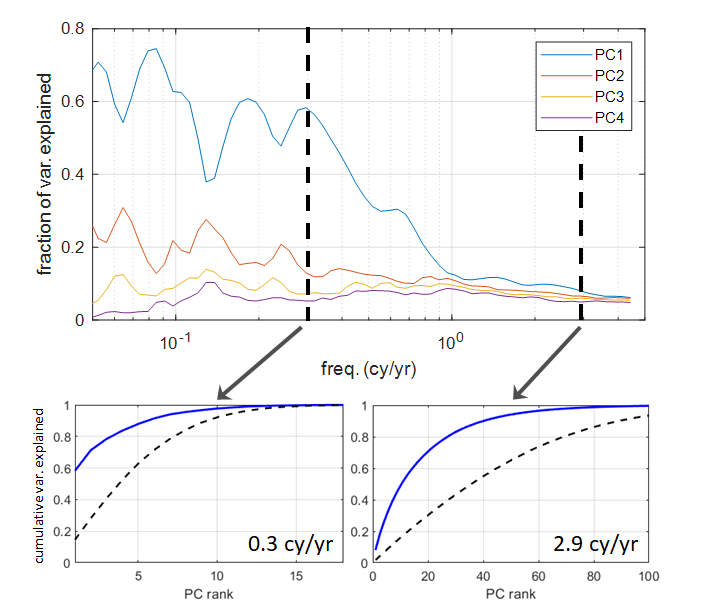}
\caption{Wavelet-based spectral PCA applied to SST monthly anomaly series over the Pacific Ocean. (Top) Fraction of the variance of the SST anomaly time series explained by the first $\mathrm{4}$ PCs of the sPCA as a function of the frequency. (Bottom, solid blue curves) cumulative variance explained as a function of the PC rank at frequencies $\mathit{f_k}\mathrm{=0.3\ {yr}^{-1}}$ and $\mathit{f_k}\mathrm{=2.9\ {yr}^{-1}}$. The dashed black curves show what would be obtained if all $\mathrm{3960}$ time series were independent random time series (colored noise).}
\label{fig9}
\end{figure} 
Figure 10 shows the first eigenvector at frequencies $\mathrm{1/3.4}$ and $\mathrm{1/5.5}\ \mathrm{yr}^{-1}$. The patterns at these frequencies are similar to the ENSO pattern that was also identified by the classical PCA. However, the phase information allows to characterize more comprehensively the general dipole structure of ENSO highlighted by the classical PCA. In particular, it shows that the phase of the response of ENSO is not the same for all extratropical and high-latitude regions, indicating non-synchronicity and propagation effects. The bottom panel of Figure 10 shows the reconstructed signal from the combination of the first wavelet PC at all frequencies between $\mathrm{1/2.6}$ and $\mathrm{1/6.8}\ \mathrm{yr}^{-1}$. This signal is consistent with the well-known historical variations of ENSO. We note that the linear trend characterizing the anthropogenic climate change does not appear in the reconstructed signal. It does not appear either in any of the first four spectral PCs at the frequencies between $\mathrm{1/20}$ and $\mathrm{1}\ \mathrm{yr}^{-1}$ (not shown). This is explained by the fact that the Morlet wavelet, being a differentiable wavelet \cite{vetterli1995wavelets,addison2017illustrated}, is blind to linear trends, i.e., the wavelet transform of a linear function of time is zero everywhere.

Contrary to what is found at low frequencies, for sub-annual frequencies it is found that there is not a single dominant PC explaining most of the system’s variability. For example, at frequency $\mathrm{3}\ \mathrm{yr}^{-1}$ ($\mathrm{4}$ months periodicity), each one of the first four PCs accounts for $\mathrm{6\%}$ to $\mathrm{8\%}$ of the system’s variability. Since several orthogonal modes of variability with similar magnitude coexist at sub-annual frequencies, the rotation of the eigenvectors with regularization of the phase through the minimal $\hat{J}(\mathbf{u}_\mathit{r})$ optimality criterion is likely to help the physical interpretation of the patterns corresponding to the extracted modes. Figure 11 shows the first two eigenvectors of the spectral PCA at frequency $\mathrm{2.5}\ \mathrm{yr}^{-1}$ with and without rotation. Before rotation, the first two modes account respectively for $\mathrm{9\%}$ and $\mathrm{7\%}$ of the system’s variability within the corresponding frequency band, and affect essentially the Northern and Southern Pacific. The latitudes between $\mathrm{20\degree}$S and $\mathrm{20\degree}$N are unaffected, except for the western Tropical Pacific around the Maritime Continent. 
\begin{figure}[!htb]
\centering
\includegraphics[width=1\textwidth]{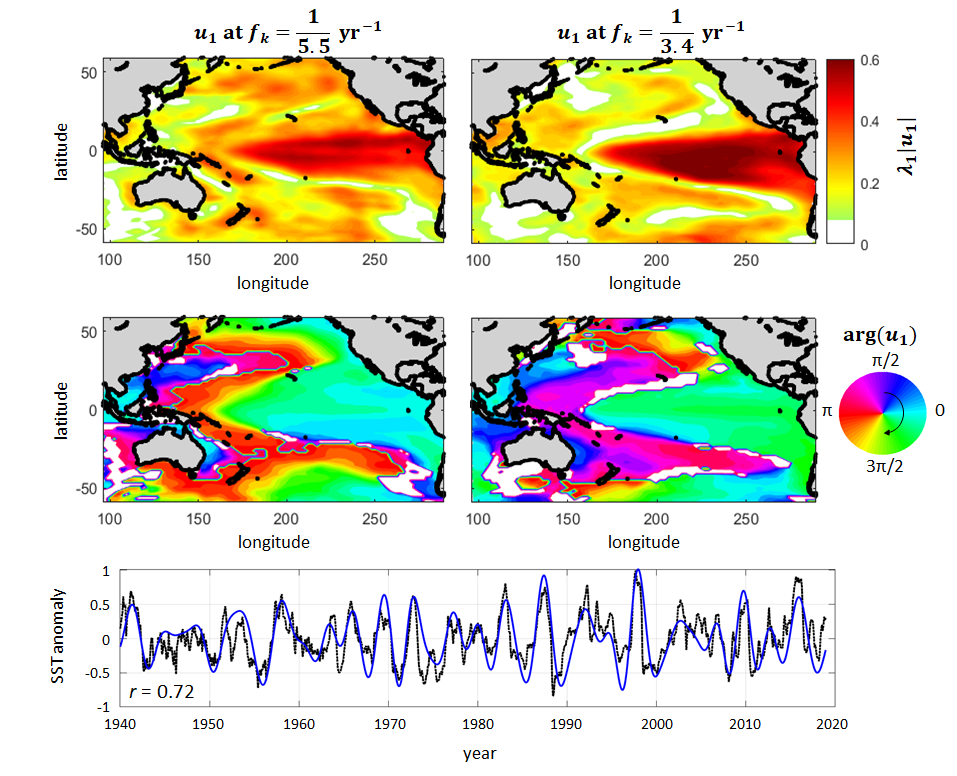}
\caption{Result of the wavelet-based spectral PCA applied to SST anomalies for low frequencies (multi-annual periods). (Top) eigenvector (phase and modulus) associated with the first PC from the spectral PCA at frequencies $\mathit{f_k}=\mathrm{\frac{1}{5.5}{yr}^{-1}}$ and  $\mathit{f_k}=\mathrm{\frac{1}{3.4}{yr}^{-1}}$. The circular arrow in the colorscale of the phase indicates the direction of propagation of the extracted waves. (Bottom, blue solid line) reconstructed signal from the combination of the first wavelet PC at every frequency between $\mathrm{1/2.6}$ and $\mathrm{1/6.8}\ \mathrm{{yr}^{-1}}$; wavelet PCs are phase adjusted before reconstruction. (Bottom, black dotted line) Niño 3.4 index (see e.g. \cite{deser2010sea}) The linear correlation between the reconstructed signal and the Niño $\mathrm{3.4}$ index is $\mathrm{0.72.}$}
\label{fig10}
\end{figure} 
\begin{figure*}[!htbp]
\centering
\includegraphics[width=\textwidth]{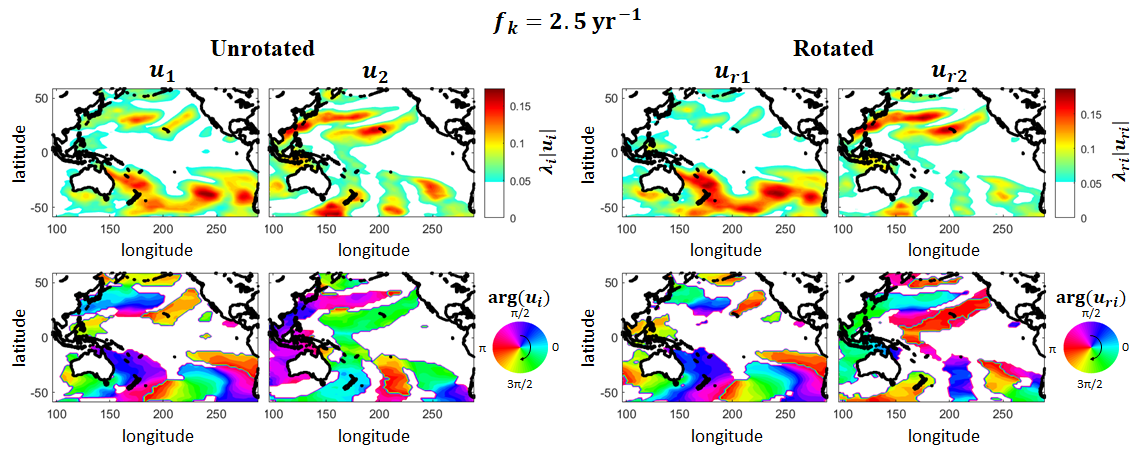}
\caption{First two eigenvectors of the wavelet-based spectral PCA at frequency $\mathit{f_k}\mathrm{=2.5\ {yr}^{-1}}$, with and without rotation. The circular arrow in the colorscale of the phase indicates the direction of propagation of the extracted waves.}
\label{fig11}
\end{figure*} 
The rotation of the first two eigenvectors successfully unmixes the two underlying modes of variability and allows for a better concentration of the first component in the Southern Pacific, and of the second one in the Northern Pacific. With regard to the first component, from the phase information we can estimate that the SST information propagates westward with a speed on the order of $\mathrm{20}$ degrees of longitude per month (i.e. around $\mathrm{0.4}$ m $\mathrm{s}^-1$). This pattern appears consistently at all frequencies between $\mathrm{2}\ \mathrm{yr}^{-1}$ and $\mathrm{3}\ \mathrm{yr}^{-1}$ (not shown) and it most probably reflects the surface signature of the South Pacific American (PSA) pattern, which is known to be the leading mode of atmospheric variability over the southern Pacific in interannual to subseasonal scales \cite{mo1998pacific,mo2001pacific,irving2016new}. In subannual scales, as in Figure 11, the PSA has been suggested to be the result of Rossby waves (westward propagation, consistent with results in Figure 11), which, when superimposed to the eastward mean circulation (the jet stream), may lead to the formation of stationary wave trains connecting the tropics and the southern mid and high latitudes in the Pacific basin \cite{irving2016new}. Accordingly, the PSA subseasonal variability has been related to the Madden-Julian oscillation \cite{madden1971detection,mo2001pacific}. The second rotated mode corresponds to a dipole in the Northern Pacific made of two bands extending respectively from the South China Sea (around $\mathrm{20\degree}$N and $\mathrm{120\degree}$E) to the latitude $\mathrm{35\degree}$N and longitude $\mathrm{190\degree}$E and from the latitude $\mathrm{20\degree}$N and longitude $\mathrm{160\degree}$E to the latitude $\mathrm{35\degree}$N and longitude $\mathrm{230\degree}$E. The opposition of phase between the two bands indicates the dipole structure (stationary wave).

The highest frequency we can theoretically have access to with monthly data is $\mathrm{6}\ \mathrm{yr}^{-1}$ (Shannon-Nyquist frequency). In practice however, because of the non-zero bandwidth of each frequency band, the highest frequency band is centered at $\mathit{f_k}=\mathrm{4.4}\ \mathrm{yr}^{-1}$. Figure 12 shows the first two (rotated) eigenvectors as obtained from the rsPCA for this frequency band. As at frequency $\mathrm{2.5}\ \mathrm{yr}^{-1}$, the first two modes are essentially extra-tropical modes, though they also affect the western tropical Pacific. The first mode appears essentially in the Western and South-Western Pacific, and specifically the southern China Sea in the proximity of the Philippines (around $\mathrm{15\degree}$N and $\mathrm{115\degree}$E) and a band in the southwestern Pacific, extending from the Coral Sea to the east of New Zealand. The latter band is collocated with the south Pacific convergence zone (SPCZ), a diagonal feature of surface wind convergence, deep convection and intense precipitation, which peaks in intensity during boreal winter \cite{widlansky2011location,haffke2013south,haffke2015diurnal,mamalakis2018multivariate}. Based on the phase information, it seems that this mode is characterized by a flow of SST information from East of New Zealand to the north, with the SST signal reaching the North-Western Pacific after a delay of $\mathrm{1-2}$ months (phase shift on the order of $\mathrm{\pi}$). This mode is consistent with an interhemispheric teleconnection recently identified by \cite{mamalakis2018new}, which has been suggested to take place during boreal fall, via the Hadley circulation. Moreover, our results show that although ENSO may be affecting the western Pacific on interannual scales (i.e., the western Pacific SSTs are part of the ENSO-like mode in Figure 10), on sub-annual scales, ENSO variability is not part of the interhemispheric teleconnection over the western Pacific in Figure 12, consistent with the finding of \cite{mamalakis2019reply}. The evolution of the magnitude of the wavelet PC associated with this mode shows that it is an intermittent mode that generally peaks once a year, in late boreal summer and winter season (not shown). This mode appears to also affect a pool off the Coast of Chile (around $\mathrm{35\degree}$S and $\mathrm{270\degree}$E) aligned in phase with the region east of New Zealand. 
\begin{figure}[!b]
\centering
\includegraphics[width=1\textwidth]{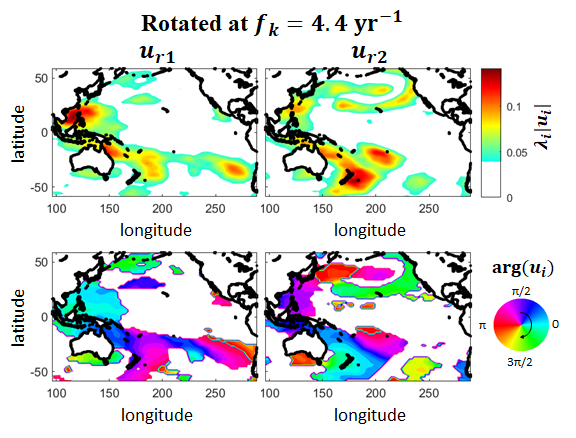}
\caption{First two eigenvectors of the rsPCA at frequency $\mathit{f_k}\mathrm{=4.4\ {yr}^{-1}}$ after rotation. The circular arrow in the colorscale of the phase indicates the direction of propagation of the extracted waves.}
\label{fig12}
\end{figure}
The second mode is geographically close to the first one, affecting the East China Sea and an area extending from the Java Sea and through the Coral Sea toward New Zealand (around $\mathrm{45\degree}$S and $\mathrm{180\degree}$E.), as well as a pool close to French Polynesia (around $\mathrm{20\degree}$S and $\mathrm{200\degree}$E) which forms a dipole (opposition of phase) with the New Zealand Area. The propagation pattern between the Coral Sea and New Zealand appears to be in opposite direction compared to the first mode. This mode may represent meridional shifts of the SPCZ on seasonal and subseasonal scales \cite{haffke2013south}. The second rotated mode at $\mathit{f_k}]mathrm{=4.4}\ \mathrm{yr}^{-1}$ also shows a dipole structure in the Northern Pacific resembling the one that was identified at $\mathit{f_k}\mathrm{=2.5}\ \mathrm{yr}^{-1}$, and thus, increasing confidence in interpretation.  
The results shown in this section clearly illustrate the advantage of the wavelet-based sPCA against the classical PCA, in separating modes that correspond to different frequencies, and in extracting propagation information. Moreover, it is demonstrated that, in the presence of several competing modes at subannual scales, rotation of the eigenvectors and optimal unmixing of the underlying climate modes via the proposed rsPCA methodology is essential for robustness and physical interpretability.

\section{Conclusion}
The need for understanding patterns of variability and change in climate signals for the purpose of predictive and diagnostic analysis, e.g., for regional prediction, untangling forced signal from internal variability, and diagnosing the performance of climate models, has never been more imperative. Classical PCA is a well-developed mathematical analysis tool which has been used extensively in climate studies. Its extension in the Fourier frequency domain, the spectral PCA (sPCA), while it can potentially better handle dynamical modes of variability due to phase information has seen limited application, arguably because of its higher implementation complexity and its sensitivity to parametric/methodological choices. We show that the implementation of spectral PCA through the continuous Morlet analytic wavelet transform offers several advantages in terms of simplicity and robustness. In the present work, particular interest is given to the phase of the eigenvectors, which contains the information for making the patterns physically interpretable. When several modes of similar amplitude exist within the same frequency band, the rotation of the eigenvectors procedure can help interpret the patterns of the emerging modes. Our proposed criterion for optimal rotation is to look for the rotated eigenvectors having the simplest phase structure which is achieved by a regularized estimation that minimizes the integral over the studied spatial area of the Laplacian of the phase of the rotated vectors. This criterion is found to be very efficient when applied to a synthetic example with two waves propagating in opposite directions.

When applied to historical SST series over the Pacific Ocean, the Morlet wavelet-based spectral PCA is able to identify the ENSO low frequency ($\mathrm{2}$ to $\mathrm{7}$ years period) mode of variability. Even without the rotation procedure, it does well in isolating the ENSO signal from other physical modes (unlike the classical PCA which mixes the ENSO and climate change modes of variability). An interesting property of the Morlet wavelet is its insensitivity to linear trends in the data (while for other spectral estimation methods de-trending of the data is often necessary). This property may be exploited to separate periodic climatic variability and localized in time “singularities” (such as volcanic events) from long term trends such as anthropogenic climate change.

At high frequencies corresponding to sub-annual variability, the spectral PCA systematically identifies several PCs explaining between $\mathrm{5\%}$ and $\mathrm{10\%}$ of the systems' variability in the corresponding frequency band. In this range of frequencies, the rotation of the eigenvectors procedure can help unmixing physical modes of variability and lead to more easily interpreted eigenvectors and PCs. The modes of variability corresponding to the first four PCs at frequencies higher than $\mathrm{2}\ \mathrm{yr}^{-1}$ can be summarized as: 1) undulating wave patterns in the Southern Pacific with a westward average propagation direction, 2) a two-way teleconnection between Southern Pacific and Western and North-Western Pacific, and 3) a dipole in the Northern Pacific. The rotation of eigenvector procedure allows in particular to separate modes affecting the same areas but with opposite propagation direction and thus identify two-way teleconnections. More generally, the phase information can allow to infer causality when the classical PCA can only determine correlations. 

The rsPCA methodology was applied here to SST data, but it can readily be applied to any spatio-temporal climatic variable. Geopotential height is one variable of particular interest which is expected to show coherent modes of variability at high frequency (weekly to monthly periodicity) \cite{madden1971detection,thompson1998arctic}. The rsPCA method holds a great potential for evaluating and comparing climate model simulations and separate climatic signal from noise by applying the method to ensembles of realizations. We note that the method may be employed using an analytic wavelet different from the Morlet wavelet. While the Morlet wavelet transform has the advantage of being relatively simply related to the Fourier transform, making the Morlet wavelet and Fourier cross-spectral matrices interpretable in a similar way, other wavelets may be more adapted for identifying modes that are not necessarily oscillatory and do not show strong periodicity. The proposed rsPCA methodology is expected to benefit from further experimentation and evaluation which may lead to potential improvements in implementation. An open question, for example, concerning this particular rsPCA and rotated PCAs in general, is how to choose the number of PCs to retain when performing rotations \cite{horel1981rotated,horel1984complex,richman1981obliquely,richman1986rotation,white1991climate}.
%
%
\appendices
\section{Rank of empirical cross spectral matrix}
\setcounter{figure}{0}
\renewcommand{\thefigure}{\thesection.\arabic{figure}}
\begin{figure}[!htbp]
\centering
\includegraphics[width=.8\textwidth]{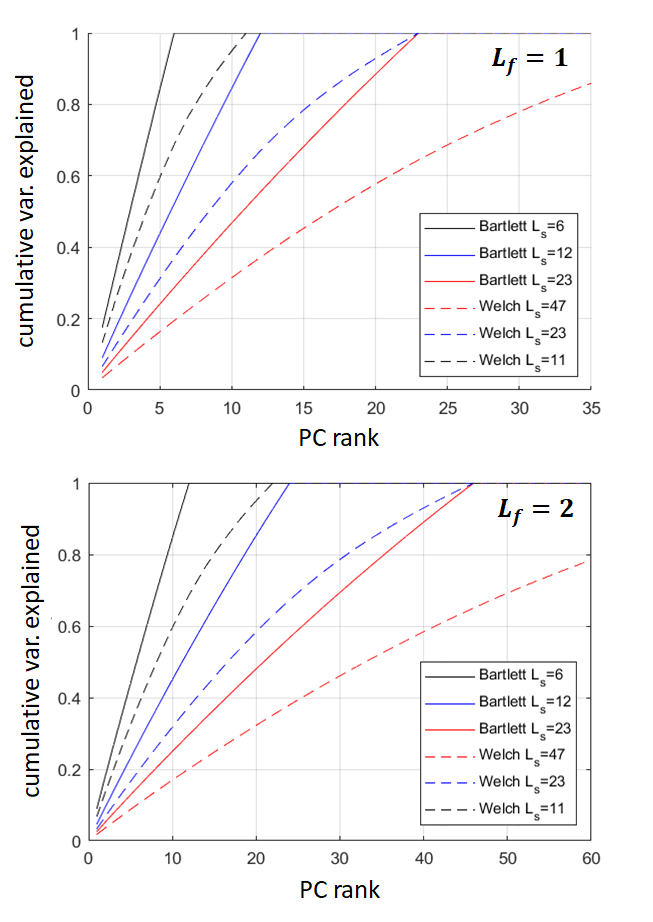}
\caption{Cumulative variance explained as a function of the PC rank for periodogram-based spectral PCAs with different windowing setups applied to a system composed of $\mathrm{3960}$ independent time series (white noise). $\mathit{L_f}$ is the number of discrete frequencies in each frequency band $\mathit{b_k}$. $\mathit{L_s}$ is the number of windows of the periodograms. For the Welch periodograms, the windowing function is a Daniel window and the overlapping factor is $\mathrm{50\%}$. Note that for the periodogram-based PCA with independent white noise time series, these results are independent of the frequency.}
\label{figA1}
\end{figure}
\setcounter{equation}{0}
\renewcommand{\theequation}{\thesection\arabic{equation}}
The empirical Fourier cross spectral matrix computed directly from the Fourier coefficients over the $\mathit{b_k}$ frequency band is:
\begin{equation}
\bar{\mathit{S}_\mathit{k}}=\frac{1}{L_sL_f}
\left[\begin{matrix}
      \sum_{f_l\in b_k}{\sum_{j = 1}^{L_s}\widehat{x_{1,j}}(f_l)\widehat{x_{1,j}}^\prime(f_l)}&\cdots&
      \sum_{f_l\in b_k}{\sum_{j = 1}^{L_s}\widehat{x_{1,j}}(f_l)\widehat{x_{N,j}}^\prime(f_l)}\\\vdots&\ddots&\vdots\\
      \sum_{f_l\in b_k}{\sum_{j = 1}^{L_s}\widehat{x_{N,j}}(f_l)\widehat{x_{1,j}}^\prime(f_l)}&\cdots&
      \sum_{f_l\in b_k}{\sum_{j = 1}^{L_s}\widehat{x_{N,j}}(f_l)\widehat{x_{N,j}}^\prime(f_l)}\\
      \end{matrix}\right]  
\end{equation}
where $\widehat{x_{1,j}}(f_l)$ is the empirical Fourier coefficient (from a discrete Fourier transform) derived from the $\mathit{j}^{th}$ sub-sample of the time series $\mathit{\mathbf{x_1}}$ at frequency $\mathit{f_l}$. $\mathit{L_f}$ is the number of number of discrete frequencies within the $\mathit{b_k}$ frequency band and $\mathit{L_s}$ is the number of sub-samples. Each sub-sample is obtained by windowing or tapering the time series. The apostrophe symbol " $\prime$ " denotes the complex conjugate operator. 
In the limit case where $\mathit{L_f}=1$ and $\mathit{L_s}=1$ (e.g. when using the discrete Fourier transform without averaging, windowing or tapering in the time domain or in the frequency domain), with  $\widehat{x_{1,1}}\left(f_l\right)=\alpha_1e^{i\theta_1}$ we obtain:
\begin{equation}
\widehat{\mathbf{S_k}}=\left[\begin{matrix}
      \begin{matrix}{\alpha_1}^2&{\ \ \alpha}_1\alpha_2e^{i(\theta_1-\theta_2)}\ \ \\\alpha_2\alpha_1e^{i(\theta_2-\theta_1)}&{\alpha_2}^2\\\end{matrix}&
      \begin{matrix}\ldots&\alpha_1\alpha_Ne^{i(\theta_1-\theta_N)}\\\ \ \ \ \ &\ \\\end{matrix}\\
      \begin{matrix}\vdots&\ \\\alpha_N\alpha_1e^{i(\theta_N-\theta_1)}&\ \ \ \ \ \ \ \ \ \ \ \ \ \ \ \ \ \ \ \ \ \ \ \ \ \ \ \ \\\end{matrix}&
      \begin{matrix}\ddots&\vdots\\\ldots&{\alpha_N}^2\\\end{matrix}\\
\end{matrix}\right]
\end{equation}
which is by construction a rank one matrix. More generally, the rank of the empirical cross-spectral matrix is at most of rank $L_s\times L_f$ since from (A1) it can be decomposed as a sum of $L_s\times L_f$ rank one matrices. The degree of independence between the sub-samples and the adjacent frequency bands also affects the decrease rate of the eigenvalues (see Figure A.1). 
\section{Fourier spectrum of the Morlet wavelet}
\setcounter{equation}{0}

The Fourier transform of the Morlet mother wavelet $\mathrm{\Psi}$ is \cite{addison2017illustrated}:
\begin{equation}
\hat{\Psi}\left(f\right)=\pi^{1/4}\sqrt2\ \ e^{-\left(2\pi f-2\pi f_o\right)^2/2}%
\end{equation}
The frequency band $\mathit{b_0}$ associated with the Morlet mother wavelet $\mathrm{\Psi}$ is therefore a Gaussian function centered at the frequency $\mathit{f_o}$. The Fourier transform of the Morlet daughter wavelet $\Psi_{t,\nu}\left(u\right)=\frac{1}{\sqrt\nu}\Psi\left(\frac{u-t}{\nu}\right)$ is:
\begin{equation}
\widehat{\Psi_{t,\nu}}\left(f\right)=\sqrt\nu\ \hat{\Psi}\left(\nu f\right)e^{-2i\pi ft}
\end{equation}
The frequency band $\mathit{b_k}$ associated with the Morlet daughter wavelet at scale $\mathit{\nu_k}$ is therefore a Gaussian function centered at the frequency $\mathit{f_k}=\ \frac{f_0}{\nu_k}$.

\section{Cone of influence and edge effects}
\setcounter{equation}{0}

When computing wavelet coefficients from finite length time series, a padding operation is needed to compute the coefficients at the beginning and at the end of the series \cite{torrence1998practical}. Possible solutions are zero-padding, repeating, or mirroring the time series. Repeating and zero padding are not recommended since they are likely to create a sharp discontinuity (particularly when the time series show a trend). An efficient solution, which is used in the present article, is to pad the series with the values corresponding to its first and last time steps, thus avoiding creating discontinuities. The most conservative option would be to not consider all the coefficients inside the "cone of influence", i.e. all the coefficients potentially affected by edge effects (e.g. by setting them to zero). However, this would reduce the length of the series of wavelet coefficients available for computing the cross-spectral matrix (particularly at coarse scales / low frequencies). 

\section{Laplacian operator and phase unwrapping}
\setcounter{equation}{0}

In image processing the discrete Laplacian operator $\hat{\mathrm{\Delta}}$ is defined as a convolution kernel of the following form:
\begin{equation}
\begin{matrix}0&-1&0\\-1&4&-1\\0&-1&0\\\end{matrix}
\end{equation}
One shall note that local unwrapping of the phase must be performed when applying the Laplacian operator in equation (17) to avoid artifacts when the phase shifts between $\mathrm{0}$ and $\mathrm{2\pi}$. It is often useful to combine the Laplacian operator with a smoothing operator to reduce sensitivity to noise. 

\section*{Acknowledgment}
The authors acknowledge support provided by the National Science Foundation (NSF) under the EAGER program (grant ECCS-1839441) and the TRIPODS+X program (grant DMS-1839336), as well as by NASA's Global Precipitation Measurement program (grant 80NSSC19K0684). L.V. was supported by a NASA Earth and Space Science Fellowship (grant 80NSSC18K1409). Upon request, the data and code that support the findings of this study can be provided by the corresponding author.

\spacingset{1}
\bibliographystyle{IEEEtran}
\bibliography{refs}

\end{document}